\RequirePackage{ifpdf}
\ifpdf 
\documentclass[pdftex]{sigma}
\else
\documentclass{sigma}
\fi

\newcommand{\al}{\alpha}
\newcommand{\pa}{\partial}
\newcommand{\ep}{\epsilon}

\newcommand{\si}{\sigma}
\newcommand{\la}{\lambda}

\newcommand{\ga}{\gamma}
\newcommand{\Om}{\Omega}
\newcommand{\om}{\omega}
\newcommand{\de}{\delta}

\newcommand{\De}{\Delta}
\newcommand{\vphi}{\varphi}

\newcommand{\rar}{\rightarrow}

\newcommand{\non}{\nonumber}

\begin{document}

\allowdisplaybreaks

\renewcommand{\thefootnote}{$\star$}

\renewcommand{\PaperNumber}{071}

\FirstPageHeading

\ShortArticleName{From Quantum $A_N$ (Calogero) to $H_4$ (Rational) Model}

\ArticleName{From Quantum $\boldsymbol{A_N}$ (Calogero)\\ to $\boldsymbol{H_4}$ (Rational) Model\footnote{This paper is a
contribution to the Special Issue ``Symmetry, Separation, Super-integrability and Special Functions~(S$^4$)''. The
full collection is available at
\href{http://www.emis.de/journals/SIGMA/S4.html}{http://www.emis.de/journals/SIGMA/S4.html}}}

\Author{Alexander~V.~TURBINER}

\AuthorNameForHeading{A.V.~Turbiner}

\Address{Instituto de Ciencias Nucleares, Universidad Nacional
Aut\'onoma de M\'exico,\\ Apartado Postal 70-543, 04510 M\'exico,
D.F., Mexico}
\Email{\href{mailto:turbiner@nucleares.unam.mx}{turbiner@nucleares.unam.mx}}

\ArticleDates{Received February 28, 2011, in f\/inal form July 12, 2011;  Published online July 18, 2011}

\Abstract{A brief and incomplete review of known integrable and (quasi)-exactly-solvable quantum models with rational (meromorphic in Cartesian coordinates) potentials is given. All of them are characterized by $(i)$~a~discrete symmetry of the Hamiltonian, $(ii)$~a~number of polynomial eigenfunctions, $(iii)$~a~factorization property for eigenfunctions, and admit $(iv)$~the separation of the radial coordinate and, hence, the existence of the 2nd order integral, $(v)$~an~algebraic form in invariants of a discrete symmetry group (in space of orbits).}


\Classification{35P99; 47A15; 47A67; 47A75}

\renewcommand{\thefootnote}{\arabic{footnote}}
\setcounter{footnote}{0}

\section{Introduction}

In this paper we will make an attempt to overview our constructive knowledge about (quasi)-exactly-solvable potentials having a form of a meromorphic function in Cartesian coordinates. All these models have a discrete group of symmetry, admit separation of variable(s), possess an (in)f\/inite set of polynomial eigenfunctions. They have an inf\/inite discrete spectrum which is linear in the quantum numbers. All of them are characterized by the presence of a hidden (Lie) algebraic structure. Each of them is a type of isospectral deformation of the isotropic harmonic oscillator.

Let us consider the Hamiltonian = the Schr\"odinger operator
\begin{gather*}
{\cal H} = -\Delta  + V(x) ,\qquad x \in {\mathbb R}^d .
\end{gather*}
The main problem of quantum mechanics is to solve the Schr\"odinger
equation
\begin{gather*}
      {\cal H} \Psi (x) = E \Psi (x)  , \qquad \Psi (x) \in L^2 \big({\mathbb R}^d\big),
\end{gather*}
f\/inding the spectrum (the energies $E$ and eigenfunctions $\Psi$).
Since the Hamiltonian is an inf\/inite-dimensional matrix, solving the Schr\"odinger equation is equivalent to diagonalizing the inf\/inite-dimensional matrix.
It is a transcendental problem: the characteristic polynomial is of inf\/inite order and it has inf\/initely-many roots. Usually, we do not know how to make such a diagonalization exactly (explicitly) but we can ask:
{\it Do models exist for which the roots $($energies$)$, some of the them or
all, can be found explicitly $($exactly$)?$}
Such models do exist and we call them {\it solvable}. If all energies are known they are called {\it exactly-solvable} (ES), if only some number of them is known we call them {\it quasi-exactly-solvable} (QES).
Surprisingly, all such models the present author familiar with, are provided by integrable systems. The Hamiltonians of these models are of the form
\begin{gather*}
 {\cal H}_{\rm ES} = -\frac{1}{2} \De + \om^2 r^2 + \frac{W(\Om)}{r^2} ,
\end{gather*}
in the exactly-solvable case and
\begin{gather*}
 {\cal H}_{\rm QES} = -\frac{1}{2} \De + \tilde\om_k^2 r^2 + \frac{W(\Om)+\Gamma}{r^2} +
 a r^6 + b r^4 ,
\end{gather*}
in the quasi-exactly-solvable case, where $\om$, $\tilde \om_k$, $\Gamma$ are parameters, $W(\Om)$ is a function on unit sphere and $r$ is the radial coordinate. In both cases there exists the integral
\begin{gather}
\label{Fi}
\mathcal{F} = \frac{1}{2}  \mathcal{L}^2 + W(\Om) ,
\end{gather}
where $\mathcal{L}$ is the angular momentum operator, due to the separation of variables in spherical coordinates.

Now we consider some examples from the ones known so far.

\section{Solvable models}

    \subsection[Case $O(N)$]{Case $\boldsymbol{O(N)}$}

The Hamiltonian reads
\begin{gather}
\label{HON}
 {\cal H}_{O(N)} = \frac{1}{2} \sum_{i=1}^{N}
 \left( -\frac{\pa^2}{\pa {x_i}^2} + \om^2 {x_i}^2 \right)
 + \frac{\nu(\nu-1)} {\sum\limits_{i=1}^{N} {x_i}^2} ,
\end{gather}
or, in spherical coordinates,
\begin{gather}
\label{HONr}
 {\cal H}_{O(N)} = -\frac{1}{2 r^N}\frac{\pa}{\pa r}\left(r^N\frac{\pa}
{\pa r}\right) + \frac{1}{2}\om^2r^2 + \frac{\mathcal{F}+\nu(\nu-1)}{r^2}  ,
\\
\label{F}
\mathcal{F} = \frac{1}{2} \mathcal{L}^2 .
\end{gather}
The Hamiltonian (\ref{HON}) is $O(N)$ symmetric. It describes a spherical-symmetric harmonic oscillator with a generalized centrifugal potential. Needless to say that the Hamiltonian~${\cal H}_{O(N)}$ and $\mathcal{F}$ commute,
\begin{gather*}
[{\cal H}_{O(N)},\mathcal{F}] = 0 .
\end{gather*}
Thus, $\mathcal{F}$ has common eigenfunctions with the Hamiltonian~${\cal H}_{O(N)}$. The spectrum can be immediately found explicitly, and all eigenfunctions are of the type
\[
  P_n\big(r^2\big) r^{\tilde\ell} Y_{\{\ell\}}(\Om) e^{-\frac{\om r^2}{2}} ,
\]
where $Y_{\{\ell\}}(\Om)$ is a $N$-dimensional spherical harmonics, $\mathcal{F} Y_{\{\ell\}}(\Om)=\gamma Y_{\{\ell\}}(\Om)$. The Hamiltonian~(\ref{HON}) describes an $N$-dimensional harmonic oscillator with generalized centrifugal term. Substituting
in (\ref{HONr}) the operator $\mathcal{F}$ by its eigenvalue $\gamma$ and gauging away
$\Psi_0=r^{\tilde\ell} e^{-\frac{\om r^2}{2}}$ we arrive at the Laguerre operator
\begin{gather}
\label{hes_ON}
    h_{O(N)} \equiv (\Psi_0)^{-1}({\cal H}_{O(N)}-E_0)\Psi_0\big|_{r^2=t}  =  -2 t\pa_t^2 +\left(2\om-1-\frac{N}{2}-\tilde\ell\right)\pa_t,
\end{gather}
where $E_0$ is the lowest energy and the parameter $\tilde\ell$ is chosen in such a way as to remove singular term $\propto \frac{1}{r^2}$ in the potential in~(\ref{HONr}). (\ref{hes_ON})~is the algebraic form of the Hamiltonian (\ref{HONr}). The gauge-rotated Hamiltonian $h_{O(N)}$ (\ref{hes_ON}) is $sl(2)$-Lie-algebraic (see below), it
has inf\/initely-many f\/inite-dimensional invariant subspaces in polynomials ${\cal P}_n$, $n=0,1,\ldots$ forming the inf\/inite f\/lag (see below), its eigenfunctions $P_n(r^2=t)$ are nothing but the associated Laguerre polynomials.

By adding to $h_{O(N)}$ (\ref{hes_ON}) the operator
\begin{gather}
\label{hqes_ON}
   \de h^{\rm (qes)}= 4 \big(a t^2 - \gamma\big) \frac{\pa}{\pa t}- 4a k t + 2\om k ,
\end{gather}
we get the operator $h_{O(N)} + \de h^{\rm (qes)}$ which has a single f\/inite-dimensional invariant subspace
\[
       {\cal P}_k = \langle t^p \,| \, 0 \leq p \leq k \rangle  ,
\]
of the dimension $(k+1)$. Hence, this operator is quasi-exactly-solvable. Making the change of variable $t=r^2$ and gauge rotation with $\tilde\Psi_0=t^{\hat \gamma}e^{-\frac{\om t}{2}-\frac{a t^2}{4}}$ we arrive at the $O(N)$-symmetric QES Hamiltonian \cite{Turbiner:1988}
\begin{gather}
\label{HON_qesr}
 {\cal H}_{O(N)} = -\frac{1}{2 r^N}\frac{\pa}{\pa r}\left(r^N\frac{\pa}
 {\pa r}\right) + a^2 r^6 + 2 a\om r^4 +
 \frac{1}{2}\tilde\om^2r^2 + \frac{{\cal F} + \Gamma}{r^2}  ,
\end{gather}
where $\hat \gamma$, $\Gamma$, $\tilde\om$ are parameters and $\gamma$ is replaced by the operator ${\cal F}$. In (\ref{HON_qesr}) a f\/inite number of the eigenfunctions is of the form
\[
  P_k\big(r^2\big) r^{2\hat \gamma} Y_{\{\ell\}}(\Om) e^{-\frac{\om r^2}{2}-\frac{a t^2}{4}} ,
\]
they can be found algebraically. It is worth noting that at $a=0$ the operator $h_{O(N)} + \de h^{\rm (qes)}$ remains exactly-solvable, it preserves the inf\/inite f\/lag of polynomials ${\cal P}$ and the emerging Hamiltonian has a form of (\ref{HONr}).

    \subsection[Case $(\mathbb{Z}_2)^{N}$]{Case $\boldsymbol{(\mathbb{Z}_2)^{N}}$}

The Hamiltonian reads
\begin{gather}
\label{HZN}
 {\cal H}_{(\mathbb{Z}_2)^N} = \frac{1}{2} \sum_{i=1}^{N}
 \left( -\frac{\pa^2}{\pa x_i^2} + \om^2 {x_i}^2 \right)
 + \frac{1}{2}\sum_{i=1}^{N} \frac{\nu_i (\nu_i-1) }{{x_i}^2}  ,
\end{gather}
or, in spherical coordinates,
\begin{gather*}
 {\cal H}_{(\mathbb{Z}_2)^N} = -\frac{1}{2 r^N}\frac{\pa}{\pa r}\left(r^N\frac{\pa}
 {\pa r}\right) + \frac{1}{2}\om^2r^2 + \frac{\mathcal{F}  +
 W_{(\mathbb{Z}_2)^N}(\Om)}{r^2} ,
\end{gather*}
where
\[
      W_{(\mathbb{Z}_2)^N}(\Om) = \frac{1}{2}\sum_{i=1}^{N} \nu_i (\nu_i-1)\left(\frac{r}{x_i}\right)^2  ,
\]
and ${\mathcal{F}}$ is given by (\ref{F}).
The Hamiltonian (\ref{HZN}) is $(\mathbb{Z}_2)^N$ symmetric. It def\/ines the so-called
Smo\-ro\-dins\-ky--Winternitz integrable system \cite{Evans:1991} which is in reality the maximally-superintegrable (there exist $(2N-1)$ integrals including the Hamiltonian) and exactly-solvable. Gauging away in~(\ref{HZN}) the ground state, $\Psi_0=\prod\limits_{i=1}^N (x_i^2)^{\frac{\nu_i}{2}} \exp \big({-}\frac{\om x_i^2}{2}\big)$, and changing variables to $t_i=x_i^2$ we arrive at the algebraic form. Also it admits QES extension.
The system described by the Hamiltonian~(\ref{HZN}) at $\nu_i=\nu$
is a particular case of the $BC_N$-rational system (see below).

\subsection[Case $A_{N-1}$]{Case $\boldsymbol{A_{N-1}}$}

\begin{figure}[!h]
\centering
\includegraphics{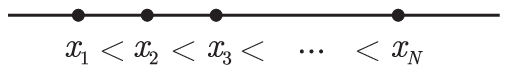}

\caption{$N$-body Calogero model.}\label{Fig1}
\end{figure}

This is the celebrated Calogero model ($A_{N-1}$ rational model) which was found in~\cite{Calogero:1969}. It descri\-bes~$N$ identical particles on a line (see Fig.~\ref{Fig1}) with singular pairwise interaction.

The Hamiltonian is
\begin{gather}
\label{HAN}
 {\cal H}_{\rm Cal} = \frac{1}{2} \sum_{i=1}^{N}
 \left( -\frac{\pa^2}{\pa x_i^2} + \om^2 x_i^2 \right)
 + \nu(\nu-1) \sum_{i>j}^{N} \frac{1}{(x_i-x_j)^2}  ,
\end{gather}
where the singular part of the potential can be written as
\begin{gather}
\label{WAN}
      \sum_{i>j}^{N}\frac{1}{(x_i-x_j)^2}  = \frac{W_{A_{N-1}}(\Om)}{r^2} ,\qquad
      W_{A_{N-1}}(\Om) = \sum_{i=1}^{N} \left(\frac{1}{\frac{x_i}{r}-\frac{x_j}{r}}\right)^2 ,
\end{gather}
Here $r$ is the radial coordinate in the space of relative coordinates
(see below for a def\/inition) and $W_{A_{N-1}}(\Om)$ is a function on the unit sphere.

Symmetry:  $S_n$ (permutations $x_i \rar x_j$)   plus   $\mathbb{Z}_2$ (all $x_i \rar -x_i$). The ground state of the Hamiltonian (\ref{HAN}) reads
\begin{gather}
\label{psi_cal}
 \Psi_{0}(x) =\prod_{i<j}|x_{i}-x_{j}|^{\nu}
e^{-\frac{\om}{2}\sum{x_i^2}}  .
\end{gather}
Let us make the gauge rotation
\[
 h_{\rm Cal} = 2\Psi_{0}^{-1}  ({\cal H}_{\rm Cal}-E_0) \Psi_{0}   ,
\]
and introduce center-of-mass variables
\[
 Y=\sum x_i  ,\qquad y_i=x_i - \frac{1}{N} Y  ,\qquad i=1,\ldots , N ,
\]
and then permutationally-symmetric, translationally-invariant variables
\[
 (x_1,x_2,\ldots, x_N) \rightarrow \big( Y, t_n (x)=\si_n (y(x))\,|\,
  n = 2,3,\ldots, N    \big)  ,
\]
where
\[
 \si_{k}(x) = \sum_{i_{1}<i_{2}<\cdots<i_{k}}
 x_{i_{1}}x_{i_{2}}\cdots x_{i_{k}} ,\qquad  \si_{k}(-x) = (-)^k \si_{k}(x)  ,
\]
are elementary symmetric polynomials, and
\[
    t_1=0  ,\qquad t_2 \sim \sum_{i<j} (x_i - x_j)^2 = r^2  ,
\]
hence, the variable $t_2$, which plays fundamental role, is def\/ined by radius in space of relative coordinates. After the center-of-mass separation, the gauge rotated Hamiltonian takes the algebraic form~\cite{RT:1995}
\begin{gather}
\label{hAN}
 h_{\rm Cal} = {\cal A}_{ij}(t) \frac{\pa^2}{\pa {t_i} \pa
{t_j} } + {\cal B}_i(t) \frac{\pa}{\pa t_i}  ,
\end{gather}
where
\begin{gather*}
  {\cal A}_{ij} =
   \frac{(N-i+1)(1-j)}{N} t_{i-1} t_{j-1} + \sum_{{l\geq}{\max (1,j-i)}} (2l-j+i)t_{i+l-1} t_{j-l-1}  ,
\\
 {\cal B}_i  =
  \frac{1}{N}(1+\nu N){(N-i+2)(N-i+1)} t_{i-2} +2\om  (i-1) t_i  .
\end{gather*}

Eigenvalues of (\ref{hAN}) are
\[
      \ep_{\{p\}}  =  2\om \sum_{i=2}^{N} (i-1)  p_i ,
\]
hence, the spectrum is linear in the quantum numbers $p_{2,3,\ldots,N}=0,1,\ldots$, it corresponds to {\it anisotropic} harmonic oscillator with frequency ratios  $1:2:3:\cdots :(N-1)$.

It is easy to check that the gauge-rotated Hamiltonian $h_{\rm Cal}$ has inf\/initely many f\/inite-dimensional invariant subspaces
\[
 {\cal P}_n^{(N-1)}  =  \langle {t_{2}}^{p_2}
 {t_{3}}^{p_3}\cdots
 {t_{N}}^{p_{N}}
 \,|\,  0 \le \sum p_i \le n \rangle .
\]
where $n=0,1,2,\ldots$. As a function of $n$ the spaces ${\cal P}_n^{(N-1)}$ form the inf\/inite f\/lag (see below).

\subsubsection[The $gl_{d+1}$-algebra acting by 1st order differential operators in ${\mathbb R}^d$]{The $\boldsymbol{gl_{d+1}}$-algebra acting by 1st order dif\/ferential operators in $\boldsymbol{{\mathbb R}^d}$}

It can be checked by the direct calculation that the $gl_{d+1}$ algebra realized by the f\/irst order dif\/ferential operators acting in ${\mathbb R}^d$ in the representation given by the Young tableaux as a row $(n,\underbrace{0,0,\ldots, 0}_{d-1} )$ has a form
\begin{gather}
 {\cal J}_i^- = \frac{\pa}{\pa t_i}, \quad
i=1,2\ldots, d  , \qquad
 {{\cal J}_{ij}}^0 =
t_i \frac{\pa}{\pa t_j}, \quad i,j=1,2,\ldots, d ,
 \qquad
{\cal J}^0 = \sum_{i=1}^{d} t_i\frac{\pa}{\pa t_i}-n ,\nonumber
 \\
 {\cal J}_i^+  =  t_i {\cal J}^0 =
t_i  \left( \sum_{j=1}^{d} t_j\frac{\pa}{\pa t_j}-n \right),
\quad i=1,2\ldots, d   . \label{gln}
\end{gather}
where $n$ is an arbitrary number. The total number of generators is $(d+1)^2$.
If $n$ takes the integer values, $n=0,1,2,\ldots$,  the f\/inite-dimensional irreps occur
\[
 {\cal P}_n^{(d)}  =  \langle {t_{1}}^{p_1}
 {t_{2}}^{p_2}\cdots
 {t_{d}}^{p_{d}}
 \,|\,  0 \le \sum p_i \le n \rangle  .
\]
It is a common invariant subspace for (\ref{gln}). The spaces ${\cal P}_n$ at $n=0,1,2,\ldots$ can be ordered
\[
{\cal P}_0 \subset  {\cal P}_1 \subset {\cal P}_2 \subset \cdots
 \subset  {\cal P}_n  \subset \cdots\subset {\cal P}  .
\]
Such a nested construction is called {\em infinite flag $($filtration$)$}
${\cal P}$. It is worth noting that the f\/lag ${\cal P}^{(d)}$ is made out of
f\/inite-dimensional irreducible representation spaces ${\cal P}_n^{(d)}$ of the
algebra $gl_{d+1}$ taken in realization~(\ref{gln}). It is evident that
{\it any operator made out of genera\-tors~\eqref{gln} has finite-dimensional invariant subspace which is finite-dimensional irreducible representation space.}

\subsubsection{Algebraic properties of the Calogero model}

It seems evident that the Hamiltonian (\ref{hAN}) has to have a representation as a second order polynomial in generators (\ref{gln}) at $d=N-1$ acting in ${\mathbb R}^{N-1}$,
\[
   h_{\rm Cal}  = {\rm  Pol}_2 \big({\cal J}_i^-  ,  {{\cal J}_{ij}}^0\big)  ,
\]
where the raising generators ${\cal J}_i^+$ are absent. Thus,
$gl(N)$ (or, strictly speaking, its maximal af\/f\/ine subalgebra) is the hidden algebra
of the $N$-body Calogero model.
Hence, $h_{\rm Cal}$ is an element of the universal enveloping algebra
       ${\cal U}_{gl(N)}$.
The eigenfunctions of the $N$-body Calogero model are
elements of the f\/lag of polynomials ${\cal P}^{(N-1)}$.
Each subspace ${\cal P}_n^{(N-1)}$ is represented by the Newton polytope (pyramid). It contains $C^{N-1}_{n+N-1}$ eigenfunctions, which is equal to the volume of the Newton polytope.

Making the gauge rotation of the integral (\ref{Fi}) with $W_{A_{N-1}}(\Om)$ given by (\ref{WAN})
\begin{gather*}
 f_{\rm Cal}  = \Psi_{0}^{-1} ({\cal F}_{\rm Cal}-F_0)\Psi_{0},
\end{gather*}
where $F_0$ is the lowest eigenvalue of the integral, ${\cal F}_{\rm Cal}\Psi_0=F_0 \Psi_0$, the integral gets the algebraic form
\[
 f_{\rm Cal} =  {f}_{ij}(t) \frac{\pa^2}{\pa {t_i} \pa
{t_j} } + {g}_i(t) \frac{\pa}{\pa t_i} ,
\]
where ${f}_{ij}$ is 2nd degree polynomial in $t$, ${f}_{2j}=0$, and ${g}_{i}$ is 1st degree polynomial in $t$, ${g}_{2}=0$. It also can be rewritten as the second degree polynomial in the $gl(N)$ generators,
\[
   f_{\rm Cal}  ={\rm Pol}_2 \big({\cal J}_i^- , {{\cal J}_{ij}}^0\big).
\]

\subsubsection[$sl(2)$-quasi-exactly-solvable generalization of the Calogero model]{$\boldsymbol{sl(2)}$-quasi-exactly-solvable generalization of the Calogero model}

By adding to $h_{\rm Cal}$ (\ref{hAN}), the operator
\begin{gather*}
   \de h^{\rm (qes)} =
   4 \big(a t_2^2 - \gamma\big) \frac{\pa}{\pa t_2}- 4a k t_2 + 2\om k ,
\end{gather*}
we get the operator $h_{\rm Cal} + \de h^{\rm (qes)}$ having f\/inite-dimensional invariant subspace
\[
       {\cal P}_k = \langle t_2^p \,| \, 0 \leq p \leq k \rangle  .
\]
By making a gauge rotation of $h_{\rm Cal} + \de h^{\rm (qes)}$
and changing of variables to Cartesian ones we arrive at the Hamiltonian~\cite{MRT}
\begin{gather*}
  {\cal H}_{\rm Cal}^{({\rm qes})} = \frac{1}{2} \sum_{i=1}^N
  \left( -\frac{\pa^2}{\pa x_i^2} + \om^2 x_i^2\right)
  + \nu(\nu-1)\sum_{j < i}^N \frac{1}{(x_i-x_j)^2}+
   \frac{2\gamma\left[\gamma-2 n(1+\nu+\nu n) + 3\right]}{{r}^2}\nonumber\\
\hphantom{{\cal H}_{\rm Cal}^{({\rm qes})} =}{}
 + a^2 {r}^6  +  2a \om {r}^4  -
 a \left[2k + 2 n(1+\nu+\nu n) - \gamma -1\right]{r}^2   .
\end{gather*}
For the Hamiltonian, $(k+1)$ eigenfunctions are of the form
\begin{gather*}
 \Psi_{k}^{({\rm qes})}(x) = \prod_{i<j}^n|x_i-x_j|^{\nu} \big(r^2\big)^{\ga} P_k \big(r^2\big)
  \exp\left[{-\frac{\om}{2} \sum_{i=1}^n x_i^2- \frac{a}{4} r^4}\right] \\
  \phantom{\Psi_{k}^{({\rm qes})}(x)}{}
   =
 (r^2)^{\ga} P_k \big(r^2\big) \exp\left(-\frac{a}{4} r^4\right) \Psi_0  ,
\end{gather*}
where $\Psi_0$ is given by (\ref{psi_cal}), $P_k$ is a polynomial of degree $k$ in  $r^2=\sum\limits_{i<j} (x_i - x_j)^2=t_2$. All remaining eigenfunctions can be represented in the same form but $P_k$'s are not polynomials anymore being functions depending on all variables~$x_i$. It is worth noting that at $a=0$ the operator $h_{\rm Cal} + \de h^{\rm (qes)}$ remains exactly-solvable, it preserves the f\/lag of polynomials ${\cal P}^{(N-1)}$ and the emerging Hamiltonian has a form of (\ref{HAN}) with the extra term $\frac{\Gamma}{r^2}$ in the potential. Its ground state eigenfunction is $(r^2)^{\ga} \Psi_0$. It is the exactly-solvable generalization of the Calogero model~(\ref{HAN}) with the Weyl group ${\rm W(A}_{N-1})$ as the discrete symmetry group,
\[
 {\cal H}_{{\rm W(A}_{N-1})}  =  {\cal H}_{\rm Cal} + \frac{\Gamma}{r^2}  .
\]

\subsection{Case: Hamiltonian reduction method}

In this method\footnote{For review and references see e.g.~\cite{OP}.} a family of integrable and exactly-solvable Hamiltonians associated with Weyl (Coxeter) symmetry was found with the Calogero model as one of its representatives. The idea of the method is beautiful and suf\/f\/icient transparent,
\begin{itemize}\itemsep=0pt
\item Take a simple group $G$,

\item Def\/ine the Laplace--Beltrami (invariant) operator on its symmetric space
      (free/harmonic oscillator motion),

\item Radial part of Laplace--Beltrami operator is the Olshanetsky--Perelomov Hamiltonian re\-le\-vant from physical point of view. The emerging Hamiltonian is the Weyl-symmetric, it can be associated with root system, it is integrable
    with integrals given by the invariant operators of higher than two orders with a property of solvability.
\end{itemize}

{\bf Rational case.}
This case appears when the coordinates of the symmetric space are introduced in such a way that the zero-curvature surface occurs. Emerging the Calogero--Moser--Sutherland--Olshanetsky--Perelomov Hamiltonian in the Cartesian coordinates has the form,
\begin{gather}
\label{HR}
 {\cal H} = \frac{1}{2}\sum_{k=1}^{N}
 \left[-\frac{\pa^{2}}{\pa x_{k}^{2}}+ \om^2 x^2_k \right] +
 \frac{1}{2}\sum_{\alpha\in R_{+}}
 \nu_{|\alpha|}(\nu_{|\al|}-1)\frac{| \al|^{2}}{(\al\cdot x)^2} ,
\end{gather}
where $R_+$ is a set of positive roots, $x$ is a position vector and $\nu_{|\al|}$ are
coupling constants (parameters) which depend on the root length.
If roots are of the same length, then $\nu_{|\al|}$ have to be equal, if all roots are
of the same length like for $A_n$, then all $\nu_{|\al|}=\nu$. In the
Hamiltonian Reduction the parameters $\nu_{|\al|}$ take a set of discrete
values, however, they can be generalized to any real value without
loosing a property of integrability as well as of solvability with the only
constraint of the existence of $L^2$-solutions of the corresponding
Schr\"odinger equation. The conf\/iguration space for (\ref{HR}) is the Weyl chamber. The ground state wave function is written explicitly,
\begin{align}
\label{psi0}
  \Psi_0 (x)   =  \prod_{\al\in R_+}
  \left|(\alpha\cdot x)\right|^{\nu_{|\al|}}e^{-\om x^2/2} .
\end{align}

The Hamiltonian (\ref{HR}) is completely-integrable: there exists a commutative algebra of integrals (including the Hamiltonian) of dimension which is equal to the dimension of the con\-f\/i\-gu\-ra\-tion space (for integrals, see Oshima~\cite{Oshima} with explicit forms of those). For each Hamiltonian~(\ref{HR}) after separation of center-of-mass coordinate (if applicable) the radial coordinate (in the space of relative coordinates)
can be also separated. It gives rise to the existence of one more integral of the second order~(\ref{Fi}). Hence, the Hamiltonian~(\ref{HR}) is super-integrable. The Hamiltonian~(\ref{HR}) is invariant with respect to the Weyl (Coxeter) group transformation, which is the discrete symmetry group of the corresponding root space.

The Hamiltonian~(\ref{HR}) has a hidden (Lie)-algebraic structure. In order to reveal it we need to
\begin{itemize}\itemsep=0pt
    \item Gauge away the ground state eigenfunction making {\it similarity transformation}
    $(\Psi_{0})^{-1} ({\cal H}-E_0) \Psi_{0}  =  h$,

    \item Consider the Hamiltonian in the space of orbits of Weyl (Coxeter)
    group by taking the {\it Weyl $($Coxeter$)$ polynomial invariants as new coordinates}, these invariants are
\[
 t_{a}^{(\Om)}(x) = \sum_{\al\in\Om} (\al \cdot x)^{a} ,
\]
where $a$'s are the {\em degrees} of the Weyl (Coxeter) group, $\Om$ is an orbit.
\end{itemize}
The invariants $t$ are def\/ined ambiguously, up to invariants of lower degrees, they depend on chosen orbit.

\subsection[Case $BC_{N}$]{Case $\boldsymbol{BC_{N}}$}

 The $BC_N$-rational model is def\/ined by the Hamiltonian,
\begin{gather}
{\cal H}_{BC_N}  =
-\frac{1}{2}\sum_{i=1}^{N}\left( \frac{\pa^2}{\pa x_i^2} -
\om^{2}x_{i}^{2}\right)
\nonumber\\
\label{HBCN}
 \phantom{{\cal H}_{BC_N}  = }{} + \nu(\nu-1) \sum_{i<j}\left[
  \frac{1}{(x_{i}-x_{j})^{2}} + \frac{1}{(x_{i}+x_{j})^{2}} \right]
 +  \frac{\nu_2(\nu_2-1)}{2}\sum_{i=1}^{N}\frac{1}{x_{i}^{2}}  ,
\end{gather}
where $\om$, $\nu$, $\nu_2$ are parameters. If $\nu=0$, the Hamiltonian~(\ref{HBCN}) is reduced to~(\ref{HZN}). The symmetry of the system is~$S_N \oplus (\mathbb{Z}_2)^{N}$ (permutations $x_i \rar x_j$ and $x_i \rar -x_i$).

The ground state function for (\ref{HBCN}) reads
\begin{gather*}
\Psi_{0} = \left[
\prod_{i<j}|x_{i}-x_{j}|^{\nu}|x_{i}+x_{j}|^{\nu}
\prod_{i=1}^{N}|x_{i}|^{\nu_{2}} \right]
e^{-\frac{\om}{2}\sum\limits_{i=1}^{N}x_{i}^{2}}  ,
\end{gather*}
(cf.~(\ref{psi0})). Making the gauge rotation
\[
h_{{BC}_N} = (\Psi_{0})^{-1}\, ({\cal H}_{BC_N}-E_0)
\Psi_{0}  ,
\]
and changing variables
\[
(x_1,x_2,\ldots, x_N) \rightarrow \big(\si_k\big(x^2\big)\,| \,
 k= 1,2,\ldots , N \big)\ ,
\]
where
\begin{gather*}
\si_{k}\big(x^2\big) = \sum_{i_{1}<i_{2}<\cdots<i_{k}}
x^2_{i_{1}}x^2_{i_{2}}\cdots x^2_{i_{k}}  ,
\qquad
   \si_1\big(x^2\big) = x_1^2+x_2^2+\cdots + x_N^2  =  r^2  ,
\end{gather*}
where $r$ is radius, we arrive at \cite{Brink:1997}
\begin{gather*}
 {h}_{BC_N}= {\cal A}_{ij}(\si) \frac{\pa^2}{\pa {\si_i} \pa
 {\si_j} } + {\cal B}_i(\si) \frac{\pa}{\pa \si_i}  ,
\end{gather*}
with coef\/f\/icients
\begin{gather*}
{\cal A}_{ij}  =  -2  \sum_{l\ge 0} (2l+1+j-i)
\si_{i-l-1} \si_{j+l}  ,
  \\
{\cal B}_i  =    \left[ 1+\nu_2 + 2\nu(N-i)\right] (N-i+1)
\si_{i-1} + 2 \om i \si_i  .
\end{gather*}
This is the algebraic form of the $BC_N$ Hamiltonian. Assuming polynomiality of the eigenfunctions we f\/ind the eigenvalues:
\[
      \ep_{n}  =  2\om \sum_{i=1}^{N} i  n_i ,
\]
hence, the spectrum is equidistant, linear in the quantum numbers and corresponds to {\it anisotropic} harmonic oscillator with frequency ratios  $1:2:3:\cdots :N$.
The Hamiltonian $h_{BC_N}$ has inf\/initely many f\/inite-dimensional invariant subspaces of the form
\[
 {\cal P}_n^{(N)}  =  \langle {\si_{1}}^{p_1}
 {\si_{2}}^{p_2}\cdots
 {\si_{N}}^{p_{N}}
 \,|\, 0 \le \sum p_i \le n \rangle  ,
\]
where $n=0,1,2,\ldots$. They naturally form the f\/lag ${\cal P}^{(N)}$. The Hamiltonian can be immediately rewritten in terms of generators~(\ref{gln})
as a polynomial of the second degree,
\[
   h_{{\rm BC}_N}  =  {\rm Pol}_2 \big({\cal J}_i^-  ,  {{\cal J}_{ij}}^0\big),
\]
where the raising generators ${\cal J}_i^+$ are absent.
Hence, $gl(N+1)$ is the hidden algebra of the~$BC_N$ rational model, the same algebra as for the $A_{N}$-rational model. The eigenfunctions of the $BC_N$-rational model are
elements of the f\/lag of polynomials ${\cal P}^{(N)}$. Each subspace ${\cal P}_n^{(N)}$ contains $C^{N}_{n+N}$ eigenfunctions (volume of the Newton polytope (pyramid) ${\cal P}_n^{(N)}$).

The $BC_N$ Hamiltonian admits 2nd order integral as result of separation of radial variable
\begin{gather*}
 {\cal H}_{BC_N}  =  - \frac{1}{2r^{N-1}}\frac{\pa}{\pa r}\left(r^{N-1}
 \frac{\pa}{\pa r}\right)  + \om^2 {r}^2 +
 \frac{1}{2r^2}\big(\underbrace{- \De_{\Om}^{(N-1)} + {\cal W}(\Om)}_{{\cal F}_{BC_N}}\big)  .
\end{gather*}
Evidently, the commutator
\[
    \left[ {\cal H}_{BC_N} ,  {\cal F}_{BC_N} \right]  = 0 .
\]
Gauge-rotated integral
\[
 f_{BC_N}  =  \Psi_{0}^{-1}  ({\cal F}_{BC_N}-F_0) \Psi_{0},
\]
where ${\cal F}_{BC_N}\Psi_0 = F_0 \Psi_0$, takes the algebraic form in $t$-coordinates,
\[
 f_{BC_N} = {f}_{ij}(t) \frac{\pa^2}{\pa {t_i} \pa
{t_j} } + {g}_i(t) \frac{\pa}{\pa t_i} ,
\]
where ${f}_{ij}$ is 2nd degree polynomial, ${f}_{1j}=0$, and ${g}_{i}$ is 1st degree polynomial, ${g}_{1}=0$,
\[
   f_{BC_N}  ={\rm Pol}_2 \big({\cal J}_i^- , {{\cal J}_{ij}}^0\big) ,
\]
in terms of the $gl(N+1)$ generators. It is worth mentioning that the commutator of $[h, f]$ vanishes only in the realization (\ref{gln}), otherwise,
\[
    [h_{BC_N} ({\cal J})  ,  f_{BC_N} ({\cal J}) ] \neq   0   .
\]

\subsubsection[$sl(2)$-quasi-exactly-solvable generalization of the $BC_N$ rational model]{$\boldsymbol{sl(2)}$-quasi-exactly-solvable generalization of the $\boldsymbol{BC_N}$ rational model}

By adding to $h_{BC_N}$ the operator
\[
   \de h^{\rm (qes)}= 4 \big(a \si_1^2 - \gamma\big) \frac{\pa}{\pa \si_1}- 4a k \si_1 + 2\om k ,
\]
which is the similar to one for the Calogero model, we get the operator $h_{BC_N} + \de h^{\rm (qes)}$ which has the f\/inite-dimensional invariant subspace
\[
       {\cal P}_k = \langle \si_1^p \,| \, 0 \leq p \leq k \rangle  .
\]
Making a gauge rotation of $h_{BC_N} + \de h^{\rm (qes)}$ and changing the variables $\si$'s back to the Cartesian ones the Hamiltonian becomes
\begin{gather*}
{\cal H}_{BC_N}^{({\rm qes})}  =
-\frac{1}{2}\sum_{i=1}^{N}\left( \frac{\pa^2}{\pa x_i^2} -
\om^{2}x_{i}^{2}\right)
 + \nu(\nu-1) \sum_{i<j}\left[
 \frac{1}{(x_{i}-x_{j})^{2}} + \frac{1}{(x_{i}+x_{j})^{2}} \right]\nonumber\\
 \hphantom{{\cal H}_{BC_N}^{({\rm qes})}  =}{}
 +   \frac{\nu_2(\nu_2-1)}{2}\sum_{i=1}^{N}\frac{1}{x_{i}^{2}}  +
   \frac{2\gamma\left[\gamma-2 N(1+2\nu(N-1) +\nu_2) + 3\right]}{{r}^2}\nonumber\\
 \hphantom{{\cal H}_{BC_N}^{({\rm qes})}  =}{}  +  a^2 {r}^6  + 2a \om {r}^4  -
 a \left[2k + 2 N(1+2\nu(N-1)+\nu_2) - \gamma -1\right]{r}^2  ,
\end{gather*}
for which $(k+1)$ eigenfunctions are of the form
\[
 \Psi_{k}^{({\rm qes})}(x) = \prod_{i<j}^n\big|x^2_i-x^2_j\big|^{\nu}
 \prod_{i=1}^{n}|x_{i}|^{\nu_2}\big(r^2\big)^{\ga} P_k \big(r^2\big)
 \ \exp\left[{-\frac{\om r^2}{2} - \frac{a}{4} r^4}\right]   ,
\]
where $P_k$ is a polynomial of degree $k$ in  $r^2=\sum\limits_{i=1}^N x_i^2$.

It is worth noting that at $a=0$ the operator $h_{BC_N} + \de h^{\rm (qes)}$ remains exactly-solvable, it preserves the f\/lag of polynomials ${\cal P}^{(N)}$ and the emerging Hamiltonian has a form of~(\ref{HBCN}) with the extra term $\frac{\Gamma}{r^2}$ in the potential. Its ground state eigenfunction is $(r^2)^{\ga} \Psi_0$. It is the exactly-solvable generalization of the $BC_N$-rational model~(\ref{HBCN}) with the Weyl group ${\rm W(BC}_{N})$ as the discrete symmetry group,
\[
 {\cal H}_{{\rm W(BC}_{N})}  =  {\cal H}_{BC_{N}} + \frac{\Gamma}{r^2}  .
\]

Now we are in a position to draw an intermediate conclusion about $A_N$ and $BC_N$ rational models.
\begin{itemize}\sloppy \itemsep=0pt
\item Both $A_N$- and $BC_N$-rational (and trigonometric) models possess
{\it algebraic} forms associated with preservation of the {\it same}
f\/lag of polynomials~${\cal P}^{(N)}$.
The f\/lag is invariant w.r.t.\ linear transformations in space of orbits
$t \mapsto t + A$. It preserves the algebraic form of Hamiltonian.

\item
Their Hamiltonians (as well as higher integrals) can be written
in the Lie-algebraic form
\[
h = {\rm Pol}_2\big({\cal J}(b\subset gl_{N+1})\big) ,
\]
where ${\rm Pol}_2$ is a polynomial of 2nd degree in generators
${\cal J}$ of the maximal af\/f\/ine subalgebra of the algebra~$b$ of the algebra
$gl_{N+1}$ in realization~(\ref{gln}).
Hence, $gl_{N+1}$ is their {\it hidden algebra}. From this viewpoint all
four models are dif\/ferent faces of a {\it single} model.

\item {\it Supersymmetric $A_N$- and $BC_N$-rational $($and trigonometric$)$
models possess {\it algebraic} forms, preserve the {\it same} flag
of $($super$)$polynomials and their {\it hidden algebra} is the super\-algebra
$gl(N+1|N)$ $($see {\rm \cite{Brink:1997})}.}
\end{itemize}

In a connection to f\/lags of polynomials we introduce a notion {\it
`characteristic vector'}.
Let us consider a f\/lag made out of ``triangular'' linear space of
polynomials
\[
 {\cal P}^{(d)}_{n, \vec f}   =   \langle x_1^{p_1}
x_2^{p_2} \cdots x_d^{p_d}\, |\, 0 \leq f_1 p_1 + f_2 p_2 +\cdots +
f_d p_d \leq n \rangle\ ,
\]
where the ``grades'' $f$'s are positive integer numbers and $n=0,1,2,\ldots$.
In lattice space ${\cal P}^{(d)}_{n, \vec f}$ def\/ines a Newton pyramid.

\begin{definition} Characteristic vector is a vector with components
$f_i$:
\[
 \vec f = (f_1, f_2, \ldots, f_d)  .
\]
From geometrical point of view $\vec f$ is normal vector to the base of the Newton pyramid. The characteristic vector for f\/lag ${\cal P}^{(d)}$ is
\[
 \vec f_0   =  \underbrace{(1,1,\ldots, 1)}_{d} .
\]
\end{definition}

  \subsection[Case $G_2$]{Case $\boldsymbol{G_2}$}

Take the Hamiltonian
\begin{gather}
{\cal H}_{G_2}   =
\frac{1}{2}\sum_{i=1}^{3}\left( -\frac{\pa^2}{\pa x_i^2}  +
\om^{2}x_{i}^{2}\right) + \nu(\nu - 1)\sum_{i<j}^3
\frac{1}{(x_{i}-x_{j})^{2}} \nonumber
\\
 \phantom{{\cal H}_{G_2}   =}{}  + 3\mu (\mu -1) \sum_{k < l,
k,l \neq m}^3 {\frac{1}{(x_{k} + x_{l}-2x_{m})^2}}  ,\label{HG2}
\end{gather}
where $\om$, $\nu$, $\mu$ are parameters.
It describes the Wolfes model of three-body interacting system~\cite{Wolfes:1974} or, in the Hamiltonian reduction nomenclature, the $G_2$-rational model.
The symmetry of the model is dihedral group~$D_6$. The ground state function is
\[
\Psi_{0} = \prod_{i<j}^3|x_i-x_j|^{\nu}
\prod^3_{k < l, \, k,l \neq m}
|x_i+x_j-2x_k|^{\mu} e^{-\frac{1}{2}\om \sum\limits_{i=1}^3 x_i^2}  .
\]
Making the gauge rotation
\[
h_{{G}_2} = (\Psi_{0})^{-1}  ({\cal H}_{G_2}-E)  \Psi_{0} ,
\]
and changing variables
\begin{gather*}
Y=\sum x_i  ,\qquad y_i=x_i - \frac{1}{3} Y  ,\quad i=1,2 , 3 ,
\qquad
(x_1,x_2,x_3) \rightarrow \big(Y,\la_1,\la_2\big),
\end{gather*}
where
\[
\la_1=-y_1^2-y_2^2-y_1 y_2 \sim -r^2   ,\qquad \la_2=[y_1 y_2(y_1+y_2)]^2 ,
\]
and separating the center-of-mass coordinate we arrive at
\begin{gather*}
h_{\rm G_2}   =   \la_1\pa^2_{\la_1\la_1}  +  6\la_2\pa^2_{\la_1\la_2}
        -  \frac{4}{3}\la_1^2\la_2\pa^2_{\la_2\la_2}
\\
\phantom{h_{G_2}   =}{}
   +  \big\{2\om\la_1+2[1+3(\mu+\nu)]\big\}\pa_{\la_1}
   + \left[6\om\la_2-\frac{4}{3}(1+2\mu)\la_1^2\right]\pa_{\la_2}  ,
\end{gather*}
which is the algebraic form of the Wolfes model. The eigenvalues of $h_{\rm G_2}$ are
\[
      \ep_{\{p\}}  =  2\om (p_1 + 3 p_2) .
\]
It coincides to the spectrum of {\it anisotropic} harmonic oscillator with frequency ratio $1:3$.

Separating the center-of-mass in (\ref{HG2}) and introducing the polar coordinates $(\varrho, \varphi)$ in the space of relative coordinates we arrive at the Hamiltonian
\begin{gather*}
 {\tilde {\cal H}}_{G_2} (\varrho,\varphi; \nu, \mu)  =  -\pa_r^2 - \frac{1}{r}\pa_r -
 \frac{1}{r^2}\pa_{\varphi}^2  + \om^2 r^2 + \frac{9\nu(\nu-1) }{r^2 \cos^2 {3 \varphi}}
 + \frac{9\mu(\mu-1) }{r^2 \sin^2 {3 \varphi}} .
\end{gather*}
It is evident that the integral of motion which appears due to separation of variables in polar coordinates (cf.~(\ref{Fi})) has the form
\begin{gather}
\label{FG2r}
  {\cal F}  =  -\pa_{\varphi}^2 + \frac{9\nu(\nu-1) }{\cos^2 {3 \varphi}}
 + \frac{9 \mu(\mu-1)}{ \sin^2 {3 \varphi}}  .
\end{gather}
It is evident that after gauge rotation with $\Psi_0$ and change of variables to $(\la_1, \la_2)$ the integral~${\cal F}$ takes algebraic form.

The Hamiltonian $h_{\rm G_2}$ has inf\/initely many f\/inite-dimensional invariant subspaces
\[
 {\cal P}_{n, (1,2)}^{(2)}  =  \langle \la_{1}^{p_1}
 \la_{2}^{p_2} \,|\,  0 \le p_1+2p_2 \le n \rangle  ,\qquad n=0,1,2,\ldots  ,
\]
hence the f\/lag ${\cal P}^{(2)}_{(1,2)}$ with the characteristic vector $\vec f =
(1,2)$ is preserved by $h_{\rm G_2}$. The eigen\-functions of $h_{\rm G_2}$ are
are elements of the f\/lag of polynomials ${\cal P}^{(2)}_{(1,2)}$.
Each subspace ${\cal P}_{n, (1,2)}^{(2)} - {\cal P}_{n-1, (1,2)}^{(2)}$ contains ${\sim}n$ eigenfunctions which is equal to length of the Newton line
${\cal L}_n n=n \langle {\la_{1}}^{p_1} {\la_{2}}^{p_2} \,|$ $p_1 + 2p_2 = n \rangle$.

A natural question to ask:
    {\it What about hidden algebra?} Namely: {\it Does algebra exist for which
    ${\cal P}_{n, (1,2)}^{(2)}$ is the space of $($irreducible$)$ representation?}
Surprisingly, this algebra exists and it is, in fact, known.

Let us consider the Lie algebra spanned by seven generators
\begin{gather}
J^1   =   \pa_t  ,
\qquad
J^2_n   =  t \pa_t  -  \frac{n}{3}   ,
 \qquad J^3_n  =  2 u\pa_u  -  \frac{n}{3}  ,
\nonumber\\
\label{gl2r}
       J^4_n   =  t^2 \pa_t    +  2 t u \pa_u   -   n t  ,
\qquad
 R_{i}  =   t^{i}\pa_u  ,\quad i=0,1,2  ,\qquad L\equiv (R_{0}, R_{1}, R_{2})  .
\end{gather}
It is non-semi-simple algebra $gl(2,{\mathbb R}) \ltimes {\cal R}^{(2)}$ (S.~Lie~\cite{Lie} at $n=0$ and A.~Gonz\'alez-Lop\'ez et al.~\cite{glko} at $n \neq 0$ (case~24)). If the parameter $n$ in (\ref{gl2r}) is a non-negative integer, it has
\[
 {\cal P}_n^{(2)}=\big(t^{p} u^{q} \,| \,0\leq (p + 2 q) \leq n\big)  ,
\]
as common (reducible) invariant subspace. By adding
\[
 T_0^{(2)}  =  u\pa_{t}^2   ,
\]
to $gl(2,{\mathbb R}) \ltimes {\cal R}^{(2)}$ (see (\ref{gl2r})),  the action on ${\cal P}_{n, (1,2)}^{(2)}$ gets irreducible.
Multiple commutators of $J^4_n$ with $T_0^{(2)}$ generate new operators acting on ${\cal P}_{n, (1,2)}^{(2)}$,
\[
    T_i^{(2)} \equiv \underbrace{[J^4,[J^4,[ \ldots J^4,T_0^{(2)}]\ldots]}_i  =
    u\pa_{t}^{2-i} J_0 (J_0+1)\cdots (J_0+i-1)   ,\quad i=0, 1,2 ,
\]
where $J_0=t\pa_t    +  2 u\pa_u   -   n$, and
all of them are of degree $2$. These new generators have a property of nilpotency,
\[
  T_i^{(2)} = 0  ,\quad i > 2  ,
\]
and commutativity:
\begin{gather}
\label{gl2t}
  \big[T_i^{(2)},T_j^{(2)}\big] = 0  ,\quad i,j=0,1,2  ,\qquad U \equiv \big(T_{0}^{(2)},T_{1}^{(2)}, T_{2}^{(2)}\big)  .
\end{gather}
(\ref{gl2r}) plus (\ref{gl2t}) span a linear space with a property of decomposition:  $g^{(2)} \doteq L \rtimes (gl_2 \oplus J_0) \ltimes U$ (see Fig.~\ref{Fig2}).

\begin{figure}[!h]
\centering\vspace{-5mm}
\begin{picture}(100,50)
\put(50,0){\vector(1,0) {18}}
\put(50,0){\vector(-1,0){18}}
\put(50,0){\vector(0,1) {15}}
\put(48,32){$g\ell_2$}
\put(30,28){\rotatebox{225}{ $\ltimes$}}
\put(68,23){\rotatebox{-45}{ $\ltimes$}}
\put(7,0){$L$}
\put(87,0){$U$}
\put(35,-17){${ P}_2{ (g\ell_2)}$}
\end{picture}
\vspace{3mm}

\caption{Triangular diagram relating the subalgebras~$L$,
$U$ and~$g\ell_2$. ${P}_2{ (g\ell_2)}$ is a polynomial of the 2nd degree in~$g\ell_2$ generators. It is a generalization of the Gauss
decomposition for semi-simple algebras.}\label{Fig2}
\end{figure}

Eventually, {\it infinite-dimensional, eleven-generated algebra $($by \eqref{gl2r} and $J_0$ plus \eqref{gl2t}, so that the eight generators are the 1st order and three generators are of the 2nd order differential operators$)$} occurs. The Hamiltonian $h_{\rm G_2}$ can be rewritten in terms of the generators (\ref{gl2r}), (\ref{gl2t}) with the absence of the highest weight generator $J^4_n$,
\begin{gather*}
  h_{\rm G_2} = \big(J^2 + 3J^3\big) J^1  -  \frac{2}{3} J^3 R_{2} + 2[3(\mu+\nu)+1] J^1
  +2\om J^2 + 3\om J^3 - \frac{4}{3} (1 + 2\mu) R_{2} ,
\end{gather*}
where $J^{2,3}=J^{2,3}_0$. Hence, $gl(2,{\mathbb R}) \ltimes {\cal R}^{(2)}$ is the hidden algebra of the Wolfes model.
\begin{enumerate}\itemsep=0pt
\item[$(i)$]  $G_2$ Hamiltonian admits two mutually-{\it non}-commuting integrals:
of 2nd order as the result of the separation of radial variable $r^2$ (see (\ref{FG2r})) and of the 6th order. If $\om=0$ the latter  integral degenerates to the 3rd order integral (the square root can be calculated in closed form).

\item[$(ii)$]  Both integrals after gauge rotation with $\Psi_0$ take in variables
$\la_{1,2}$ the algebraic form. Both preserve the same f\/lag ${\cal P}^{(2)}_{(1,2)}$.

\item[$(iii)$] Both integrals can be rewritten in term of generators of the algebra
$g^{(2)}$: integral of 2nd order in terms of $gl(2,{\mathbb R}) \ltimes {\cal R}^{(2)}$ generators only and while one of the 6th order contains generators from $\mathfrak{L}$ as well~\cite{TTW:2009}.
\end{enumerate}

    \subsubsection[$sl(2)$-quasi-exactly-solvable generalization]{$\boldsymbol{sl(2)}$-quasi-exactly-solvable generalization}

By adding to $h_{G_2}$, the operator (the same as for the Calogero and the $BC_N$ models)
\[
   \de h^{\rm (qes)}= 4 (a \la_1^2 - \gamma) \frac{\pa}{\pa \la_1}- 4a k \la_1 + 2\om k  ,
\]
we get the operator $h_{G_2} + \de h^{\rm (qes)}$ having single f\/inite-dimensional invariant subspace
\[
       {\cal P}_k = \langle \la_1^p \, | \, 0 \leq p \leq k \rangle  .
\]
Making a gauge rotation of $h_{G_2} + \de h^{\rm (qes)}$, changing of variables $(Y,\la_{1,2})$ back to the Cartesian coordinates and adding the center-of-mass the Hamiltonian becomes
\begin{gather*}
  {\cal H}_{G_2}^{\rm (qes)}  =
  -\frac{1}{2}\sum_{i=1}^{3}\left( \frac{\pa^2}{\pa x_i^2}-\om^{2}x_{i}^{2}\right)  +
  \nu(\nu-1) \sum_{i<j}^3 \frac{1}{(x_{i}-x_{j})^{2}} \nonumber\\
\phantom{{\cal H}_{G_2}^{\rm (qes)}  =}{}
  +
 3\mu(\mu-1) \sum_{i<l,\, i,l \neq m}^{3} \frac{1}{(x_{i} + x_{l}-2x_{m})^2}
 +  \frac{4\gamma(\gamma+3\mu+3\nu)}{{r}^2}
\nonumber\\
\phantom{{\cal H}_{G_2}^{\rm (qes)}  =}{}   +
 a^2 {r}^6 + 2a \om {r}^4 +
 2a\left[2k - 3(\mu+\nu) - 2(\gamma +1) \right]{r}^2  ,
\end{gather*}
for which $(k+1)$ eigenfunctions are of the form
\begin{gather*}
 \Psi_k^{({\rm qes})}=
 \prod_{i<j}^3 |x_i-x_j|^{\nu}
 \prod^3_{i<j;i,j\neq p}|x_i+x_j-2x_p|^{\mu}\big(r^2\big)^{\ga} P_k \big(r^2\big)
 \exp\left[{-\frac{\om}{2}\sum_{i=1}^3 x_i^2 - \frac{a}{4} r^4}\right]  ,
\end{gather*}
where $P_k$ is a polynomial of degree $k$ in  $r^2$.

It is worth noting that at $a=0$ the operator $h_{G_2} + \de h^{\rm (qes)}$ remains exactly-solvable, it preserves the f\/lag of polynomials ${\cal P}^{(2)}_{(1,2)}$ and the emerging Hamiltonian has a form of (\ref{HG2}) with the extra term $\frac{\Gamma}{r^2}$ in the potential. Its ground state eigenfunction is $(r^2)^{\ga} \Psi_0$. It is the exactly-solvable generalization of the $G_2$-rational model (\ref{HG2}) with the Weyl group $\rm W(G_{2})$ as the discrete symmetry group,
\[
 {\cal H}_{\rm W(G_{2})}  =  {\cal H}_{G_{2}} + \frac{\Gamma}{r^2}  .
\]

  \subsection[Cases $F_4$ and $E_{6,7,8}$]{Cases $\boldsymbol{F_4}$ and $\boldsymbol{E_{6,7,8}}$}

In some details these four cases are described in~\cite{BLT:2005}.

  \subsection[Case $I_2 (k)$]{Case $\boldsymbol{I_2 (k)}$}

In some details this case is described in~\cite{TTW:2009}. It is worth noting that although the Hamiltonian Reduction nomenclature is assigned to this case the parameter $k$ takes
any real value. Discrete symmetry group~$D_{2k}$ of the Hamiltonian appears for integer~$k$.

  \subsection[Case $H_{3}$]{Case $\boldsymbol{H_{3}}$}

The $H_{3}$ rational Hamiltonian reads
\begin{gather}
    \mathcal{H}_{H_3}= \frac{1}{2}\sum_{k=1}^{3}\left[-\frac{\pa^{2}}{\pa
    x_{k}^{2}}+\om^{2}x_{k}^{2}+\frac{\nu(\nu-1)}{x_{k}^{2}}\right]  \non \\
\phantom{\mathcal{H}_{H_3}=}{}
+2\nu(\nu-1)\sum_{\{i,j,k\}} \sum_{\mu_{1,2}=0,1}\frac{1}{[x_{i}+(-1)^{\mu_1}
    \varphi_{+}x_{j}+(-1)^{\mu_2}\varphi_{-}x_{k}]^{2}}  ,\label{H3}
\end{gather}
where $\{i,j,k\}=\{1,2,3\}$ and all even permutations, $\om$, $\nu$ are parameters and
\[
    \varphi_{\pm}=\frac{1\pm\sqrt{5}}{2}  ,
\]
the golden ratio and its algebraic conjugate.
Symmetry of the Hamiltonian~(\ref{H3}) is the ${\rm H_3}$ Coxeter group (the full symmetry group of the icosahedron). It has the order 120. In total, the Hamiltonian~(\ref{H3}) is symmetric with respect to the transformation
\begin{gather*}
    x_i   \longleftrightarrow  x_j  , \qquad
    \vphi_+   \longleftrightarrow  \vphi_-  .
\end{gather*}
The ground state is given by
\[
    \Psi_{0}=\Delta_{1}^{\nu}\Delta_{2}^{\nu}\exp\left(-\frac{\om}{2}
    \sum_{k=1}^{3}x_k^2\right)  ,\qquad E_{0}=\frac{3}{2}\om(1+10\nu)  ,
\]
where
\begin{gather*}
    \Delta_1 =\prod_{k=1}^{3}x_k  ,\qquad
    \Delta_2 =\prod_{\{i,j,k\}}\prod_{\mu_{1,2}=0,1}\left[x_i+(-1)^{\mu_1}
    \vphi_+x_j+(-1)^{\mu_2}\vphi_-x_k\right] .
\end{gather*}
Making the gauge rotation
\[
    h_{H_3}=-2(\Psi_{0})^{-1}(\mathcal{H}_{H_3}-E_{0})(\Psi_{0}) ,
\]
we arrive at new spectral problem
\[
    h_{H_3}\phi(x)=-2\ep\phi(x) .
\]
After changing variables $(x_{1,2,3} \rar \tau_{1,2,3})$:
\begin{gather*}
    \tau_1= x_1^2+x_2^2+x_3^2 = r^2\ , \\
\tau_2= -\frac{3}{10} \big(x_1^6+x_2^6+x_3^6\big)+\frac{3}{10}(2-5\vphi_+)     \big(x_1^2x_2^4+x_2^2x_3^4+x_3^2x_1^4\big) \\
 \phantom{\tau_2=}{} +\frac{3}{10}(2-5\vphi_-)\big(x_1^2x_3^4+x_2^2x_1^4+x_3^2x_2^4\big)-\frac{39}{5}  ,\\
    \tau_3= \frac{2}{125}\big(x_1^{10}+x_2^{10}+x_3^{10}\big)+\frac{2}{25}(1+5\vphi_-)
    \big(x_1^8x_2^2+x_2^8x_3^2+x_3^8x_1^2\big)\\
    \phantom{\tau_3=}{}
 +\frac{2}{25}(1+5\vphi_+)\big(x_1^8x_3^2+x_2^8x_1^2+x_3^8x_2^2\big)
 +\frac{4}{25}(1-5\vphi_-)\big(x_1^6x_2^4+x_2^6x_3^4+x_3^6x_1^4\big)\\
 \phantom{\tau_3=}{}
 +\frac{4}{25}(1-5\vphi_+)\big(x_1^6x_3^4+x_2^6x_1^4+x_3^6x_2^4\big)
 -\frac{112}{25}\big(x_1^6x_2^2x_3^2+x_2^6x_3^2x_1^2+x_3^6x_1^2x_2^2\big)\\
 \phantom{\tau_3=}{}
+\frac{212}{25}\big(x_1^2x_2^4x_3^4+x_2^2x_3^4x_1^4+x_3^2x_1^4x_2^4\big)  ,
\end{gather*}
in the gauge-rotated Hamiltonian, it emerges in the algebraic form \cite{H3}
\[
    h_{H_3}=\sum_{i,j=1}^{3}A_{ij}\frac{\pa^2}{\pa\tau_i\pa\tau_j}+
    \sum_{j=1}^{3}B_j\frac{\pa}{\pa\tau_j}  ,
\]
where
\begin{gather*}
    A_{11}=4\tau_{1}  ,\qquad A_{12}=12\tau_{2}   ,\qquad A_{13}=20\tau_{3}  ,
\\
  A_{22} = -\frac{48}{5}\tau_{1}^2\tau_2+\frac{45}{2}\tau_3  ,\qquad
  A_{23} = \frac{16}{15}\tau_1\tau_2^2-24\tau_1^2\tau_3  ,\qquad
  A_{33} = -\frac{64}{3}\tau_1\tau_2\tau_3+\frac{128}{45}\tau_2^3  ,
\\
    B_{1} = 6+60\nu-4\om\tau_1   ,\qquad
    B_{2} = -\frac{48}{5}(1+5\nu)\tau_{1}^{2}-12\om\tau_2  ,
\\
    B_{3} = -\frac{64}{15}(2+5\nu)\tau_1\tau_2-20\om\tau_3  ,
\end{gather*}
which is amazingly simple comparing the quite complicated and lengthy form of the original Hamiltonian~(\ref{H3}). The Hamiltonian $h_{H_3}$ preserves inf\/initely-many spaces
\[
    \mathcal{P}_{n}^{(1,2,3)}=\langle
    \tau_{1}^{n_{1}}\tau_{2}^{n_{2}}\tau_{3}^{n_{3}}\,|\,0\leq
    n_{1}+2n_{2}+3n_{3}\leq n\rangle  ,\qquad n\in\mathbb{N}  ,
\]
with characteristic vector is $(1,2,3)$,
they form an inf\/inite f\/lag. The spectrum of $h_{H_3}$ is given by
\[
    \ep_{p_1,p_2,p_3}=2\om(p_1+3p_2+5p_3)  ,\qquad
    p_i=0,1,2,\ldots  ,
\]
with degeneracy $p_1+3p_2+5p_3=\text{integer}$. It corresponds to the anisotropic harmonic oscillator with frequency ratios $1:3:5$.
Eigenfunctions $\phi_{n,i}$ of $h_{H_3}$ are elements of $\mathcal{P}_n^{(1,2,3)}$,
The number of eigenfunctions in $\mathcal{P}_n^{(1,2,3)}$ is maximal possible~-- it is equal to dimension of $\mathcal{P}_n^{(1,2,3)}$.

The space $\mathcal{P}^{(1,2,3)}_{n}$ is f\/inite-dimensional representation space of a Lie algebra of dif\/ferential operators which we call the $h^{(3)}$ algebra. It is inf\/inite-dimensional but f\/initely generated algebra of dif\/ferential operators with 30 generating elements of $1^{\text{st}}$ (14), $2^{\text{nd}}$ (10) and $3^{\text{rd}}$ (5) orders, respectively, plus one of zeroth order. They span $5 + 5$ Abelian (conjugated) subalgebras of lowering and raising generators $L$ and $U$\footnote{These subalgebras can be divided into pairs. In every pair the elements of dif\/ferent subalgebras are related via a certain conjugation.} and one of the Cartan type algebra $B$ (for details see~\cite{H3}). The algebra $B=g\ell_2 \oplus I_2$, where $I_2$ is two-dimensional ideal. The Hamiltonian~$h_{H_3}$ can be rewritten in terms of the generators of the $h^{(3)}$-algebra.

By adding to $h_{H_3}$ (\ref{H3}) the operator (\ref{hqes_ON}) in the variable $\tau_1$
(of the same type as for the Calogero, $BC_N$ and $G_2$ models) we get the operator $h_{H_3} + \de h^{\rm (qes)}$ which has the f\/inite-dimensional invariant subspace
\[
       {\cal P}_k = \langle {\tau_1}^p \,| \, 0 \leq p \leq k \rangle  ,
\]
of the dimension $(k+1)$. Hence, this operator is quasi-exactly-solvable. Making the gauge rotation of this operator and changing variables $\tau$ back to Cartesian ones we arrive at the quasi-exactly-solvable Hamiltonian of a similar type as for the Calogero, $BC_N$ and $G_2$ models \cite{H3}. By adding to $h_{H_3}$ (\ref{H3}) the operator $4 \gamma \frac{\pa}{\pa\tau_1}$ we preserve the property of exact-solvability. This operator preserves the f\/lag $\mathcal{P}^{(1,2,3)}$ and the emerging Hamiltonian has a form of (\ref{H3}) with the extra term $\frac{\Gamma}{r^2}$ in the potential. Its ground state eigenfunction is $(r^2)^{\ga} \Psi_0$. It is the exactly-solvable generalization of the $H_3$-rational model (\ref{H3}) with the Coxeter group ${\rm H_{3}}$ as the discrete symmetry group,
\[
 {\cal H}_{\rm H_{3}}  =  {\cal H}_{H_{3}} + \frac{\Gamma}{r^2}  .
\]

  \subsection[Case $H_{4}$]{Case $\boldsymbol{H_{4}}$}

The $H_{4}$ rational Hamiltonian reads
\begin{gather}
    \mathcal{H}_{H_4}= \frac{1}{2}\sum_{k=1}^{4}\left[-\frac{\pa^{2}}{\pa
    x_k^2}+\omega^{2}x_{k}^{2}+\frac{\nu(\nu-1)}{x_{k}^{2}}\right]   \non \\
\phantom{\mathcal{H}_{H_4}=}{}
+2\nu(\nu-1)\sum_{\mu_{2,3,4}=0,1}\frac{1}
    {[x_1+(-1)^{\mu_2}x_2+(-1)^{\mu_3}x_3+(-1)^{\mu_4}x_4]^2}
\nonumber\\
\phantom{\mathcal{H}_{H_4}=}{}
 +   2\nu(\nu-1)\sum_{\{i,j,k,l\}} \sum_{\mu_{1,2}=0,1}\frac{1}
    {[x_{i}+(-1)^{\mu_1}\varphi_{+}x_{j}+(-1)^{\mu_2}\varphi_{-}x_{k}+0\cdot x_{l}]^{2}}  ,\label{H4}
\end{gather}
where $\{i,j,k,l\}=\{1,2,3,4\}$ and all even permutations, $\om$, $\nu$ are parameters and
\[
    \varphi_{\pm}=\frac{1\pm\sqrt{5}}{2}   ,
\]
the golden ratio and it algebraic conjugate.
Symmetry of the Hamiltonian (\ref{H4}) is the ${\rm H_4}$ Coxeter group
(the symmetry group of the {\em $600$-cell}). It has order 14400.
In total, the Hamiltonian~(\ref{H4}) is symmetric with respect to the transformation
\[
    x_i \longleftrightarrow x_j  ,\qquad
    \vphi_+ \longleftrightarrow \vphi_-  .
\]
The ground state function and its eigenvalue are
\[
    \Psi_{0}=\Delta_{1}^{\nu}\Delta_{2}^{\nu}\Delta_{3}^{\nu}\exp\left(-\frac{\om}{2}
    \sum_{k=1}^{4}x_{k}^{2}\right)  ,\qquad E_{0}=2\om(1+30\nu)  ,
\]
where
\begin{gather*}
    \De_1 =\prod_{k=1}^{4}x_{k}  ,\qquad
    \De_2 =\prod_{\mu_{2,3,4}=0,1}
    \big[x_1+(-1)^{\mu_2}x_2+(-1)^{\mu_3}x_3+(-1)^{\mu_4}x_4\big]  ,\\
    \De_3 =\prod_{\{i,j,k,l\}} \prod_{\mu_{1,2}=0,1}
    \big[x_{i}+(-1)^{\mu_1}\varphi_{+}x_{j}+(-1)^{\mu_2}\varphi_{-}x_{k}+0\cdot x_l\big] .
\end{gather*}
Making a gauge rotation of the Hamiltonian
\[
    h_{H_4}=-2(\Psi_{0})^{-1}(\mathcal{H}_{H_4}-E_{0})(\Psi_{0})  ,
\]
and introducing new variables $\tau_{1,2,3,4}$, which are invariant with respect to the ${\rm H_4}$ Coxeter group, in a form of polynomials in $x$ of degrees~2, 12, 20, 30 (degrees of ${\rm H_4}$), we arrive at the Hamiltonian in the algebraic form~\cite{H4}
\begin{gather}
\label{hh4}
    h_{H_4}=\sum_{i,j=1}^{4}A_{ij}\frac{\pa^2}{\pa\tau_i\pa\tau_j}+
    \sum_{j=1}^{4}B_j\frac{\pa}{\pa\tau_j}  ,
\end{gather}
where
\begin{gather*}
    A_{11}= 4  \tau_{1}    ,\qquad
    A_{12}= 24  \tau_{2}   ,\qquad
    A_{13}= 40  \tau_{3}    ,\qquad
    A_{14}= 60  \tau_{4}  ,
\\
    A_{22}= 88  \tau_1\tau_3 + 8  \tau_1^5\tau_2   ,\qquad
    A_{23}= -4  \tau_1^3\tau_2^2 + 24  \tau_1^5\tau_3 - 8  \tau_4 ,
\\
    A_{24} = 10  \tau_1^2\tau_2^3 + 60 \tau_1^4\tau_2\tau_3 + 40 \tau_1^5\tau_4 - 600  \tau_3^2  ,
\qquad
    A_{33} = -\frac{38}{3} \tau_1\tau_2^3 + 28  \tau_1^3\tau_2\tau_3 - \frac{8}{3}  \tau_1^4\tau_4   ,
\\
    A_{34} = 210  \tau_1^2\tau_2^2\tau_3 + 60  \tau_1^3\tau_2\tau_4 - 180  \tau_1^4\tau_3^2 + 30  \tau_2^4  ,
\\
    A_{44}= -2175  \tau_1\tau_2^3\tau_3 - 450\tau_1^2\tau_2^2\tau_4 - 1350 \tau_1^3\tau_2\tau_3^2 - 600 \tau_1^4\tau_3\tau_4 ,
\\
    B_{1} = 8(1+30\nu) - 4\om\tau_1   , \qquad
    B_{2} = 12(1+10\nu)\ \tau_1^5 - 24\om\tau_2  ,
\\
    B_{3} = 20(1+6\nu)  \tau_1^3\tau_2 - 40\om\tau_3   ,\qquad
    B_{4} = 15(1-30\nu)  \tau_1^2\tau_2^2 - 450(1+2\nu)
    \tau_1^4\tau_3-60\om\tau_4  ,
\end{gather*}
which is amazingly simple comparing the very complicated and lengthy form of the original Hamiltonian (\ref{H4}).
It is easy to check that the algebraic operator $h_{H_4}$ preserves inf\/initely-many f\/inite-dimensional invariant subspaces
\[
    \mathcal{P}_{n}^{(1,5,8,12)}=\langle
    \tau_{1}^{n_{1}}\tau_{2}^{n_{2}}\tau_{3}^{n_{3}}\tau_{4}^{n_{4}}\, |\, 0\leq
    n_{1}+5n_{2}+8n_{3}+12n_{4}\leq n\rangle  ,\qquad n\in\mathbb{N}  ,
\]
all of them with the same characteristic vector $(1, 5, 8, 12)$, they form the inf\/inite f\/lag. The spectrum of the Hamiltonian~$h_{H_4}$~(\ref{hh4}) has a form
\[
    \ep_{k_1,k_2,k_3,k_4}=2\om(k_1+6k_2+10k_3+15k_4)  ,\qquad
    k_i=0,1,2,\ldots ,
\]
with degeneracy $k_1+6k_2+10k_3+15k_4=\text{integer}$. It corresponds to the anisotropic harmonic oscillator with frequency ratios $1:6:10:15$.
Eigenfunctions $\phi_{n,i}$ of $h_{H_4}$ are elements of~$\mathcal{P}_n^{(1,5,8,12)}$. The number of eigenfunctions in $\mathcal{P}_n^{(1,5,8,12)}$ is equal to dimension of $\mathcal{P}_n^{(1,5,8,12)}$.

The space $\mathcal{P}^{(1,5,8,12)}_{n}$ is a f\/inite-dimensional representation space of a Lie algebra of dif\/ferential operators which we call the $h^{(4)}$ algebra. It is inf\/inite-dimensional but f\/initely generated algebra of dif\/ferential operators, with 183 generating elements of $1^{\text{st}}$ (54), $2^{\text{nd}}$ (24), $3^{\text{rd}}$ (18), $4^{\text{rd}}$ (18), $5^{\text{rd}}$ (28), $6^{\text{rd}}$ (5), $8^{\text{rd}}$ (20), $9^{\text{rd}}$ (1), $12^{\text{rd}}$ (14) orders, respectively, plus one of zeroth order. They span $26 + 26$ Abelian (conjugated) subalgebras of lowering and raising generators $L$ and $U$\footnote{These subalgebras can be divided into pairs. In every pair the elements of dif\/ferent subalgebras are related via a certain conjugation.} and one of the Cartan type algebra $B$ (for details see~\cite{H4M}). The algebra $B=g\ell_2 \oplus I_3$, where $I_3$ is three-dimensional ideal. The Hamiltonian $h_{H_4}$ can be rewritten in terms of the generators of the $h^{(4)}$-algebra.

By adding to $h_{H_4}$ (\ref{H4}) the operator (\ref{hqes_ON}) in the variable $\tau_1$
(of the same type as for the Calogero, $BC_N$ and $G_2$, $H_3$ models) we get the operator $h_{H_4} + \de h^{\rm (qes)}$, which has the f\/inite-dimensional invariant subspace
\[
       {\cal P}_k = \langle {\tau_1}^p \,|\, 0 \leq p \leq k \rangle  ,
\]
of dimension $(k+1)$. Hence, this operator is quasi-exactly-solvable. Making the gauge rotation of this operator and changing variables $\tau$ back to Cartesian ones we arrive at the quasi-exactly-solvable Hamiltonian of a similar type as for the Calogero, $BC_N$ and $G_2$ models \cite{H4}. By adding to $h_{H_4}$ (\ref{H4}) the operator $4 \gamma \frac{\pa}{\pa\tau_1}$ we preserve the property of exact-solvability. This operator preserves the f\/lag $\mathcal{P}^{(1,5,8,12)}$ and the emerging Hamiltonian has a form of (\ref{H4}) with the extra term $\frac{\Gamma}{r^2}$ in the potential. Its ground state eigenfunction is $(r^2)^{\ga} \Psi_0$. It is the exactly-solvable generalization of the $H_4$-rational model~(\ref{H4}) with the Coxeter group ${\rm H_{4}}$ as the discrete symmetry group,
\[
 {\cal H}_{\rm H_{4}}  =  {\cal H}_{H_{4}} + \frac{\Gamma}{r^2}  .
\]

\section{Conclusions}

 \begin{itemize}\itemsep=-1pt
\item
For rational Hamiltonians for all classical $A_N$, $BC_N$ and exceptional root spaces $G_2$, $F_4$, $E_{6,7,8}$ (also trigonometric) and non-crystallographic $H_{3,4}$, $I_{2}(k)$ there exists an algebraic form after gauging away the ground state eigenfunction, and changing variables to symmetric (invariant) variables. Their eigenfunctions are polynomials in these variables. They are orthogonal with the squared ground state eigenfunction as the weight factor.

\item
Their hidden algebras are $gl(N+1)$ for the case of classical $A_N$, $BC_N$ and
{\it new} inf\/inite-dimensional but f\/inite-generated algebras of dif\/ferential operators for all other cases. All these algebras have f\/inite-dimensional invariant subspace(s) in polynomials.

\begin{figure}[t]
\centering\vspace{-1mm}
\begin{picture}(110,110)
\put(55,50){\vector(-1,0){45}} \put(55,50){\vector(1,0){45}}
\put(55,50){\vector(0,1){45}} \put(105,45){$U$}
\put(0,45){$L$} \put(43,35){$P_{p}(B)$} \put(50,100){$B$}
\put(75,80){\scalebox{1.7}{\rotatebox{315}{$\ltimes$}}}
\put(15,87){\scalebox{1.7}{\rotatebox{225}{$\ltimes$}}}
\end{picture}\vspace{-13mm}
\caption{Triangular diagram relating the subalgebras $L$,
$U$ and $B$. ${P}_p(B)$ is a polynomial of the $p$th degree in $B$ generators. It is a generalization of the Gauss decomposition for semi-simple algebras where $p=1$.}
\label{Fig3}
\end{figure}
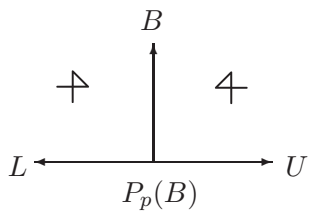

\item
The generating elements of any such hidden algebra can be grouped into an even number of (conjugated) Abelian algebras $L_i$, $U_i$ and one Lie algebra $B$. They obey a (genera\-li\-zed) Gauss decomposition (see Fig.~\ref{Fig3}). A description of all these algebras will be given elsewhere.
\end{itemize}

There are two solvable potentials in $1D$ in $[0,\infty)$ which can be generalized
to $D$.

\subsection{General view (exact-solvability)}

ES-case generalization:

\[
\om^2 r^2 + \frac{\gamma}{r^2} \quad \rar \quad \om^2 r^2 + \frac{\gamma(\Om)}{r^2}
\]
(a generalization by replacing the symmetry $O(N)$ by its discrete subgroup of symmetry given by Weyl (Coxeter) group).

\subsection{General view (quasi-exact-solvability)}

QES-case generalization:
\[
\om^2 r^2 + \frac{\gamma}{r^2} + a r^6 + b r^4 \quad \rar \quad \tilde\om^2 r^2 + \frac{\tilde\gamma (\Om)}{r^2} + a r^6 + b r^4
\]
(a generalization by replacing the symmetry $O(N)$ by its discrete subgroup of symmetry given by Weyl (Coxeter) group).

In conclusion I must emphasize that the algebraic nature of all above-considered systems was revealed when:
\begin{center}
{\it The invariants of the discrete group of symmetry of a system\\ are taken as variables $($space of orbits$)$.}
\end{center}

\subsection*{Acknowledgements}

This paper is based on the talk given
to honor Professor Willard Miller, who was always an example for me of
love and dedication for science.

\pdfbookmark[1]{References}{ref}

\LastPageEnding

\end{document}